\documentclass[aps,twocolumn,showpacs,superscriptaddress,pra,10pt]{revtex4-1}
\usepackage{graphicx,braket,amsmath,amsfonts,hyperref,color}

\begin{document}

\author{Mario Motta}
\affiliation{Division of Chemistry and Chemical Engineering, California Institute of Technology, 
Pasadena, CA 91125, USA}

\author{Shiwei Zhang}
\affiliation{Department of Physics, College of William and Mary, Williamsburg, Virginia 23187-8795, USA}

\title{Calculation of interatomic forces and 
optimization of molecular geometry
with auxiliary-field quantum Monte Carlo}

\begin{abstract}

We propose an algorithm for accurate, systematic and scalable computation of interatomic 
forces within the auxiliary-field Quantum Monte Carlo (AFQMC) method.
The algorithm relies on the Hellman-Fenyman theorem, 
and incorporates Pulay corrections in the presence of atomic orbital basis sets.
We benchmark the method for small molecules
by comparing the computed forces with  the derivatives of the AFQMC 
potential energy surface, and by direct comparison with other quantum chemistry methods.
We then perform geometry optimizations using the steepest descent algorithm in larger molecules. 
With realistic basis sets, we obtain
equilibrium geometries in agreement, within statistical error bars, with experimental values.
The increase in computational cost  
for computing forces in this approach is only a small prefactor over that 
of calculating the total energy. 
This paves the way for a general and efficient approach 
for geometry optimization and molecular dynamics within AFQMC.
\end{abstract}

\maketitle


Calculating interatomic forces in molecules is important for a quantitative 
understanding of their physical 
properties.
Forces are the basic ingredient in a variety of fundamental studies 
including 
molecular dynamics simulations 
\cite{Car_PRL55_1985}, 
optimization of molecular geometries 
\cite{Pulay_MP17_1969,Eckert_JCC18_1997,Pulay_WIRES4_2014}, 
computation of vibrational properties 
\cite{Pulay_WIRES4_2014} 
and reaction path following 
\cite{Eckert_TCA100_1998}.

Despite the incredible success 
within the framework of density functional theory (DFT) 
\cite{Reveles_JCC25_2004},
much effort has been devoted to developing many-body schemes that can describe more accurately 
physical and chemical situations where DFT is less reliable, for instance in cases of large 
electronic correlations and long-range dispersive forces 
\cite{Busch_JCP94_1991,Azhary_JCP108_1998,Rauhut_PCCP3_2001}.
Quantum Monte Carlo (QMC) methods are one of the 
promising alternative many-body approaches, 
which treat electronic correlations by stochastically sampling of 
correlated wavefunctions.

The auxiliary-field quantum Monte Carlo (AFQMC) method \cite{Zhang_PRL90_2003}, in particular, 
has seen rapid development and applications to a wide variety of quantum chemistry and condensed-matter systems \cite{AlSaidi_JCP124_2006,
Suewattana_PRB75_2007,Purwanto_JCP128_2008,
Purwanto_JCP130_2009,Purwanto_JCP142_2015,Purwanto_JCP144_2016,Kwee_PRL100_2008,
Purwanto_PRB80_2009,Ma_PRL114_2015,Shee2017}, 
yielding state-of-the-art, benchmark-quality results \cite{Motta_PRX_2017}
for the energies of ground and excited states.
It is intrinsically 
parallel, having tremendous capacity to take advantage of petascale (and bebyond) computing resources \cite{Esler_JP125_2008}. 
The computational cost scales as the third or fourth power of the system size,
 offering the potential to treat large many-electron systems. 

The calculation of forces with QMC methods has been a long standing problem
\cite{Zong_PRE58_1998,Casalegno_JCP118_2003,Lee_JCP122_2005,Sorella_JCP133_2010}.
Despite significant recent development
\cite{Assaraf_JCP113_2000,Assaraf_JCP119_2003,Filippi_PRB61_2000,
Casalegno_JCP118_2003,Chiesa_PRL94_2005, Moroni_JCTC10_2014},
computing forces reliably and efficiently beyond variational Monte Carlo (VMC) methods
has remained a major challenge. 

In this paper, we present 
an algorithm to directly calculate interatomic forces within AFQMC.
As the AFQMC algorithm works by sampling a non-orthogonal Slater determinant space, its formalism
allows the adaptation of several 
key ingredients from other electronic structure or quantum chemistry methods. Combining this with recent advances in the back-propagation technique \cite{Motta_JCTC_2017}, we achieve an efficient approach for computing atomic forces 
within AFQMC.
In this paper we describe the algorithm, and demonstrate it by full QMC geometry optimization in molecules 
using a simple steepest descent approach. 
Internal consistency of the algorithm is verified by measuring 
the agreement between computed forces and the gradients of the potential energy surface.
We benchmark the accuracy of the algorithm by comparison with available high-level quantum chemistry 
and/or experimental results.


The AFQMC method \cite{Zhang_PRL90_2003,Zhang_notes_2013,Motta_WIRES_2017} reaches 
the many-body ground state $\Psi_0$ of a Hamiltonian $\hat{H}$ by an iterative process, 
$
|\Psi_0\rangle \propto \lim_{n\to\infty} \exp(- n \, \Delta\tau \hat{H}) \, |\Psi_T\rangle
$,
where $\Delta\tau$ is a small parameter and, for convenience, we take the initial state $\Psi_T$,
which must be non-orthogonal to $\Psi_0$,
to be a single Slater determinant. The many-body propagator is written as 
\begin{equation}
e^{-\Delta\tau \hat{H}} = 
\int d{\bf{x}} \, p({\bf{x}}) \, \hat{B}({\bf{x}}) \, ,
\end{equation}
where $\hat{B}({\bf{x}})$ is an independent-particle propagator 
that depends on the multi-dimensional 
vector ${\bf{x}}$, and $p({\bf{x}})$ is a probability distribution function \cite{Zhang_PRL90_2003}.
AFQMC thus represents the many-body wave function in the iteration as an ensemble of Slater determinants, 
\begin{equation}
\label{eq:psi-PI-form}
\begin{split}
|\Psi^{(n)}\rangle 
& = e^{- n \, \Delta\tau \hat{H}} |\Psi_T\rangle
\propto \int d\Psi \, \lambda_n(\Psi) |\Psi\rangle \, , \\
\lambda_n(\Psi) &= \int \prod_{l=0}^{n-1} d{\bf{x}}_l \,\, p({\bf{x}}_l) 
\,\, \delta \left(\Psi, \prod_{l=0}^{n-1} \hat{B}({\bf{x}}_l) \Psi_T \right) \, , 
\end{split}
\end{equation}
where  $\int d\Psi$ denotes integration over the manifold of Slater determinants $\Psi$ \cite{Fahy_PRB43_1991}.

The ground-state expectation value of 
an operator $\hat{A}$ can be obtained as
\begin{equation}
\label{eq:esta}
\frac{
\langle \Psi^{(m)} | \hat{A} | \Psi^{(n)} \rangle
}{
\langle \Psi^{(m)} | \Psi^{(n)} \rangle 
}
= \frac
{\int d\Phi d\Psi \, \rho_{m,n}(\Phi,\Psi) \, W(\Phi,\Psi) \, A_{loc}(\Phi,\Psi) }
{\int d\Phi d\Psi \, \rho_{m,n}(\Phi,\Psi) \, W(\Phi,\Psi) } \, ,
\end{equation}
in the limit of large $n$ and $m$, 
where $\rho_{m,n}(\Phi,\Psi) = \mu^*_m(\Phi) \lambda_n(\Psi)$.
Formally $\mu_m$ in the distribution $\rho$ should be the same as  $ \lambda_m$;
however, we have used a different symbol to emphasize that in AFQMC the paths leading to it
are obtained by  the back-propagation algorithm \cite{Purwanto_PRE70_2004,Motta_JCTC_2017}.
Operationally this can be thought of as generating $N_w$ Monte Carlo (MC) path configurations of auxiliary fields,
$\{{\bf{x}}_0, {\bf{x}}_1, \cdots, {\bf{x}}_{n-1}; {\bf{x}}_{n}, \cdots,  {\bf{x}}_{n+m-1}\}$, in the 
stochastic sampling of $\langle \Psi^{(m)} | \Psi^{(n)} \rangle=\langle \Psi_T | \Psi^{(n+m)} \rangle$, and then back-propagate $\langle \Psi_T | $ for $m$ steps 
along the path to obtain $\langle \Psi^{(m)} |$,
\begin{equation}
\mu_m(\Phi) = 
\int \prod_{l=n}^{n+m-1} d{\bf{x}}_l \,\, p({\bf{x}}_l) \,\, 
\delta \left(\Phi, \prod_{l=n+m-1}^n \hat{B}({\bf{x}}_l)^\dagger \Psi_T \right) \,.
\end{equation}
In Eq.~\eqref{eq:esta},  $|\Psi\rangle$ and $\langle \Phi|$ in
the overlap 
$W(\Phi,\Psi) = \langle \Phi | \Psi \rangle$ 
and the local expectation
$A_{loc}(\Phi,\Psi) = \langle \Phi | \hat{A} | \Psi \rangle/\langle \Phi | \Psi \rangle$
are defined on the path at ``time-slices'' $(n-1)$ and $n$, respectively.
The expectation is evaluated over the MC samples (labeled by $w$) as
\begin{equation}
\frac{
\langle \Psi_0 | \hat{A} | \Psi_0 \rangle 
}{
\langle \Psi_0 | \Psi_0 \rangle 
}
\simeq 
\frac{ \sum_{w=1}^{N_w} W(\Phi_w,\Psi_w) \, A_{loc}(\Phi_w,\Psi_w) }
{ \sum_{w=1}^{N_w} W(\Phi_w,\Psi_w) } \, .
\end{equation}

Because the propagator $\hat{B}({\bf{x}})$ contains stochastically fluctuating fields, 
the MC sampling will lead to negative (indeed complex) overlaps $W$, which will 
cause the variance of this estimator to grow exponentially with the projection times, 
$n$ and $m$. Control of this phase problem is achieved by the introduction 
of an importance sampling transformation and a generalized gauge condition 
to constrain the random walks \cite{Zhang_PRL90_2003}.

This framework eliminates the phase problem, at the cost of modifying 
the distribution $\rho_{m,n}$. The constraint introduces a bias in $\lambda_n$.
For $\mu_m$ an additional subtlety arises because the constraint on the path is imposed in 
the time-reversed direction \cite{Motta_WIRES_2017}.
Previous studies in a variety of systems have shown that the bias from the constraint tends 
to be small, in both models \cite{LeBlanc_PRX5_2015} and realistic materials 
\cite{Motta_PRX_2017}, 
making AFQMC of the most accurate many-body approaches for general interacting fermion 
systems.


To address the problem of computing atomic forces, let us consider a molecule comprising $N_n$ 
ions, whose spatial positions ${\bf{R}} = ({\bf{R}}_1 \dots {\bf{R}}_{N_n})$ define a molecular geometry. 
Given a molecular geometry ${\bf{R}}$, ground-state expectation values of 
physical observables are given by
\begin{equation}
\label{eq:field}
A({\bf{R}}) = 
\frac{  \langle \Psi_0 ({\bf{R}}) | \hat{A}({\bf{R}}) | \Psi_0 ({\bf{R}}) \rangle  }
{\langle \Psi_0 ({\bf{R}}) | \Psi_0 ({\bf{R}}) \rangle} 
\equiv \mathcal{A}[ \rho_{m,n} ;{\bf{R}}] \, .
\end{equation}
For $\hat{A}({\bf{R}})=\hat{H}({\bf{R}})$, Eq.~\eqref{eq:field} gives the 
potential energy surface (PES), 
$E({\bf{R}})$. The gradient of \eqref{eq:field} is in general given by
\begin{equation}
\label{eq:gradient-0}
\frac{\partial A}{\partial R_\alpha} = 
\frac{\partial \rho_{m,n}}{\partial R_\alpha} 
\frac{\partial \mathcal{A}}{\partial \rho_{m,n}}
+
\frac{\partial \mathcal{A}}{\partial R_\alpha}\,.
\end{equation}
In the case of the PES,  the first two terms vanish for sufficiently large $m$ and $n$.
However, the expression does not reduce to 
Hellmann-Feynman theorem \cite{Hellmann_book_1937,Feynman_PR56_1939}, because 
$\Psi$ and $\Phi$ are represented by incomplete atom-centered basis sets which depend 
on ${\bf{R}}$.
Performing the partial derivative in Eq.~\eqref{eq:esta}, one obtains
\begin{equation}
\begin{split}
\label{eq:gradient}
&F^\alpha({\bf{R}}) = \frac{\partial E}{\partial R_\alpha}({\bf{R}}) = \\ 
= \, &\frac
{\int d\Phi d\Psi \, \rho_{m,n}(\Phi,\Psi) \, W(\Phi,\Psi;{\bf{R}}) \, F^\alpha_{loc}(\Phi,\Psi;{\bf{R}}) }
{\int d\Phi d\Psi \, \rho_{m,n}(\Phi,\Psi) \, W(\Phi,\Psi;{\bf{R}}) }\,,
\end{split}
\end{equation}
with 
\begin{equation}
\label{eq:gradient_integrand}
\begin{split}
&F^\alpha_{loc}(\Phi,\Psi;{\bf{R}}) =  \frac{\partial E_{loc}}{\partial R_ \alpha} (\Phi,\Psi;{\bf{R}}) \\
+ \, & \frac{\partial_{R_ \alpha} W( \Phi , \Psi; {\bf{R}}) }{W( \Phi , \Psi; {\bf{R}})} 
\big(E_{loc}(\Phi,\Psi;{\bf{R}}) - E({\bf{R}}) \big) \, . 
\end{split}
\end{equation}
The terms in Eq.~\eqref{eq:gradient_integrand} are detailed in the Appendix, 
and encompass a Pulay's correction \cite{Pulay_MP17_1969} to take into account the explicit dependence of the 
one-electron Green's function on atomic coordinates ${\bf{R}}$.

Ground-state expectation values of the form in Eq.~\eqref{eq:gradient}
can be computed accurately 
with the back-propagation algorithm, originally formulated for lattice models of correlated electrons 
\cite{Zhang_PRB_1997}, generalized to  weakly correlated systems with complex $\hat B({\bf x})$
\cite{Purwanto_PRE70_2004}, and recently adapted to molecular 
systems with a ``path restoration" (BP-PRes) technique \cite{Motta_JCTC_2017}.
The computational cost of evaluating $F^\alpha_{loc}(\Phi,\Psi;{\bf{R}})$  
is similar to 
that of computing 
the local energy.
Additional speedup can be achieved by exploiting the local nature of the atomic forces and the sparsity of the Hamiltonian
matrix elements.
In larger molecules and solids, the ability 
to evaluate all forces simultaneously (as opposed to separate total energy calculations for each force component) 
is crucial, and paves the way for geometry optimizations.

We note that the formalism and estimator detailed above hold for any other method operating in an
overcomplete manifold of non-orthogonal Slater determinants. Differences arise in the generation of 
the distributions $\mu$ and $\lambda$ which, as discussed above, need not be differentiated
to compute the atomic forces.


\begin{figure}[ht!]
\includegraphics[width=0.45\textwidth]{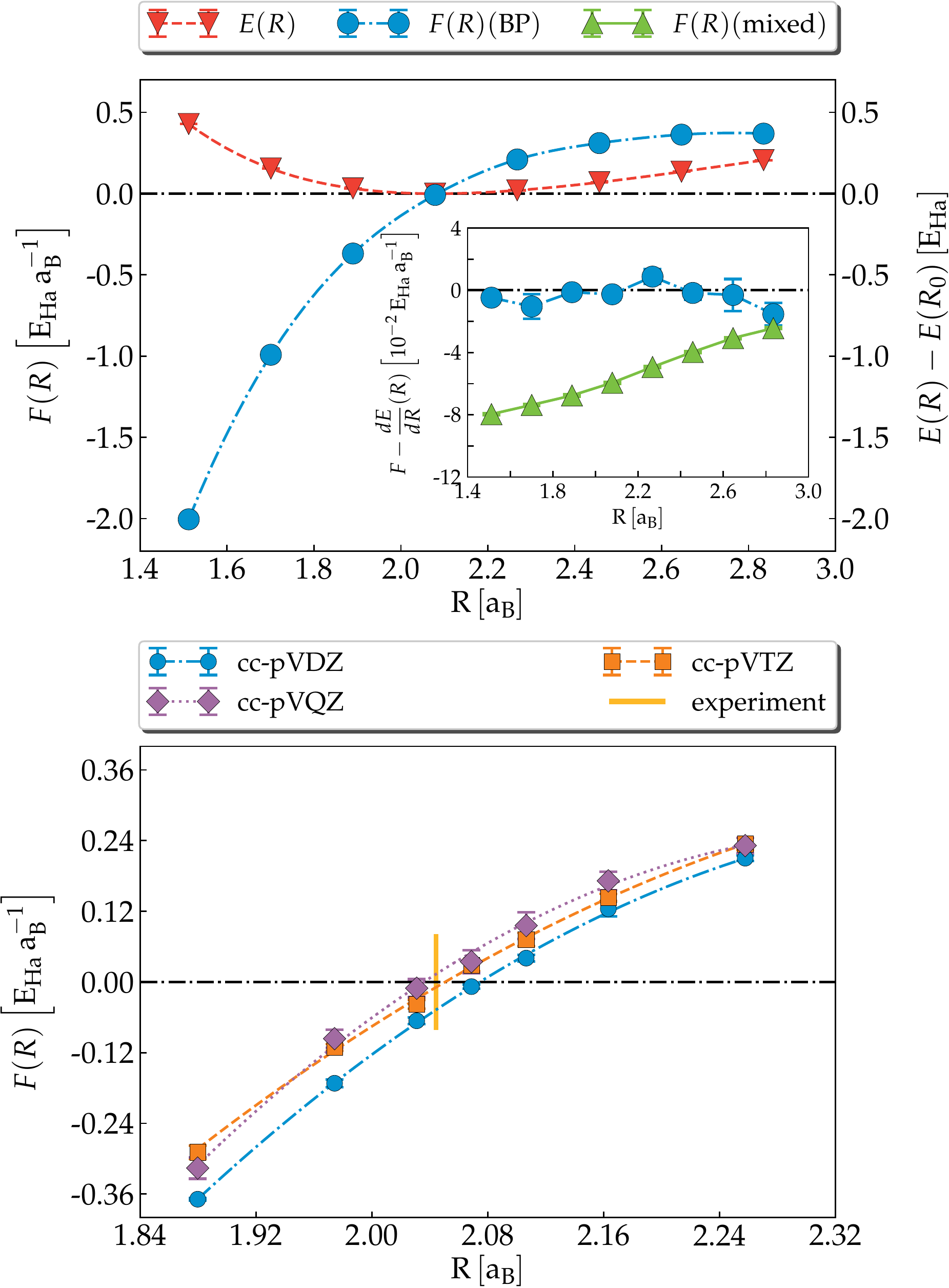}
\caption{(color online) Top: AFQMC potential energy surface $E(R)$ (red triangles) 
and computed force $F(R)$ (blue points) of CH$_4$ as function of the CH bondlength $R$
(cc-pVDZ basis).
The minimum of the energy $E(R)$ is attained when $F(R)=0$, as expected.
Inset: comparison between computed force $F(R)$ using back-propagation (blue points) and mixed 
estimator, i.e. $m=0$ in Eq.~\eqref{eq:esta} (green triangles), and 
numerical finite difference of $dE(R)/dR$, taken as reference.
Bottom: AFQMC computed force $F(R)$ of CH$_4$, using cc-pVxZ basis sets. 
}
\label{fig:ch4}
\end{figure}

In Fig.~\ref{fig:ch4} we assess the internal consistency and accuracy of our method 
using the symmetric stretching of the CH bond in CH$_4$ as a test case. 
The computed gradients, across the entire bondlength range,
 are in agreement with results from explicit (numerical) derivative of the PES within statistical 
uncertainties. 
As the inset in the upper panel illustrates, it is crucial to have BP in order for Eq.~\eqref{eq:gradient} to
 achieve an accurate representation of Eq.~\eqref{eq:gradient-0}.
Performing calculations with increasingly large basis sets, we observe convergence of the 
equilibrium bondlength to the complete basis set (CBS) limit 
($R=2.0844(32)$, $2.0605(70)$, $2.0458(62)\,\mathrm{a_B}$ at cc-pVDZ, TZ, QZ level 
respectively), and good agreement with the experimental equilibrium bondlength 
($R=2.0541(19)\,\mathrm{a_B}$  \cite{CCCBDB2016}).

\begin{figure}
\includegraphics[width=0.45\textwidth]{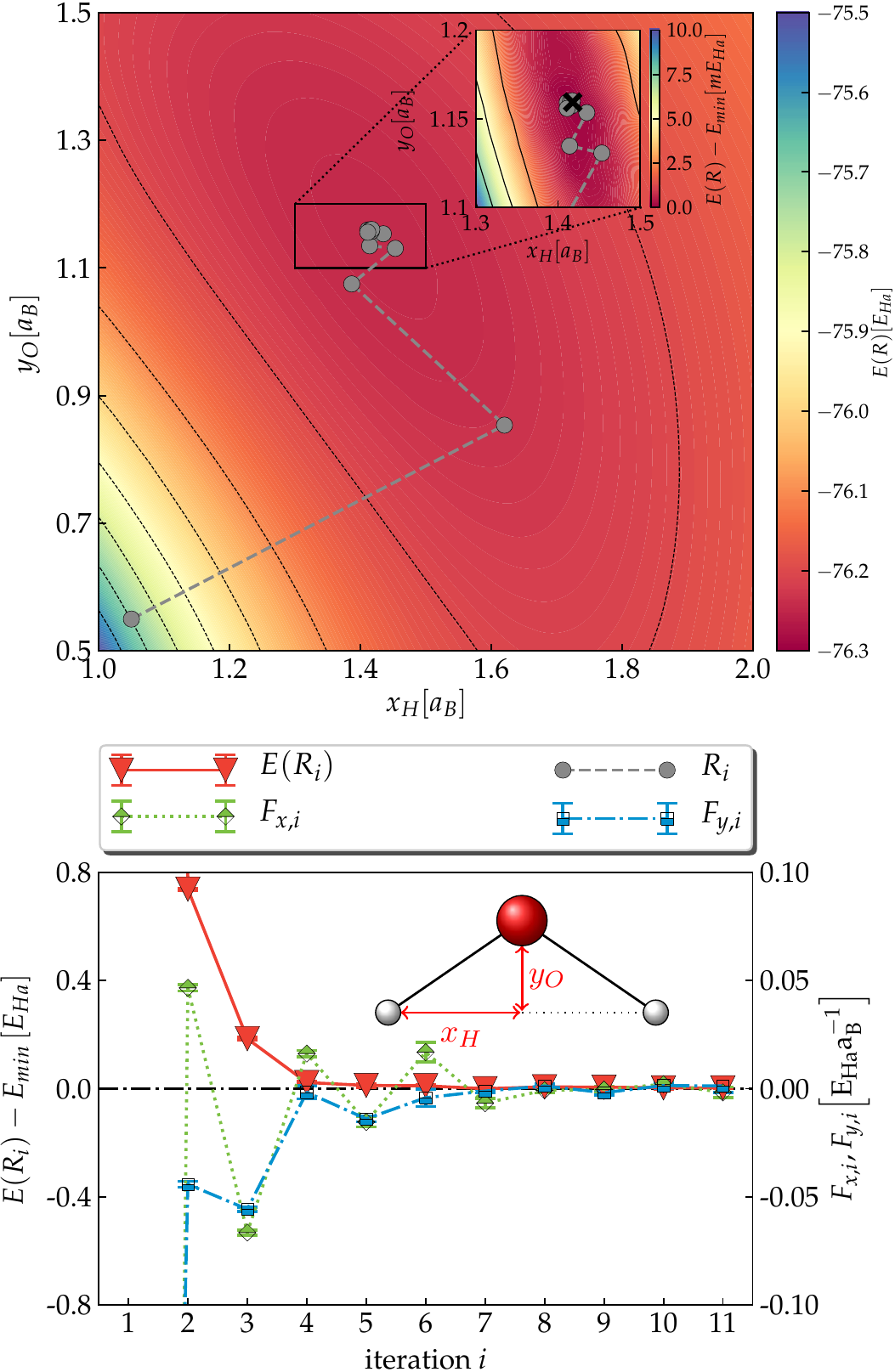}
\caption{(color online) Top: The PES, 
$E({\bf{R}})$, computed on a dense mesh from AFQMC   (color plot) 
and steps (gray points) of a steepest-descent geometry optimization for H$_2$O, at cc-pVDZ level.
Inset: magnified view around the equilibrium geometry. 
The location of the global minimum of PES is indicated with a cross ($x_H=1.4179(12) \,\mathrm{a_B}$, 
$y_O = 1.1593(12) \, \mathrm{a_B}$).
Bottom: convergence of the AFQMC energy $E({\bf{R}}_i)$ and extinction of the residual forces 
$F_{x,i} = \partial E({\bf{R}}_i)/\partial x_H$ (green diamonds) and 
$F_{y,i} = \partial E({\bf{R}}_i)/\partial y_O$ (blue squares)
 during the steepest-descent geometry 
optimization. The relevant geometry parameters $x_H$ and $y_O$ are sketched.
}
\label{fig:h2o}
\end{figure}

As a further test, 
we perform a geometry optimization in Fig.~\ref{fig:h2o} for the H$_2$O molecule 
using the steepest-descent algorithm.
The computed equilibrium geometry is in agreement with the global minimum of the 
PES, 
obtained from AFQMC calculations of the ground-state energy on a dense mesh 
of points. 
In Table~\ref{tab:h2o} we 
list the optimized geometries for  different basis sets, together with those from the 
 coupled-cluster CCSD(T) for comparison and benchmark. 
The statistical error bars on the QMC geometry 
are obtained by averaging over configurations of $\{{\bf{R}}_i\}$ 
with converged and statistically compatible energies, for example, at the cc-pVDZ level, 
$i=6 \dots 11$ (see bottom panel in Fig.~\ref{fig:h2o}).
From Table~\ref{tab:h2o}, convergence to the CBS limit is seen at the cc-pVQZ level, where the AFQMC results are in good agreement with experiment \cite{CCCBDB2016}.

\begin{table}
\begin{tabular}{cccc}
\hline\hline
basis & method & $x_H$ [$\mathrm{a_B}$] & $y_O$ [$\mathrm{a_B}$] \\
\hline
cc-pVDZ & CCSD(T)    & 1.418     & 1.149 \\
               & AFQMC      & 1.414(2) & 1.158(1) \\
cc-pVTZ & CCSD(T)     & 1.424     & 1.118 \\
               & AFQMC      & 1.426(3) & 1.121(4) \\
cc-pVQZ & CCSD(T)    & 1.439     & 1.109 \\
               & AFQMC      & 1.425(5) & 1.098(4) \\
               & experiment  & 1.431     & 1.108 \\
\hline\hline
\end{tabular}
\caption{Equilibrium 
geometries of H$_2$O computed from AFQMC with several basis sets, compared to the corresponding
CCSD(T) result and experimental  \cite{CCCBDB2016} equilibrium 
geometries. The geometry is defined in terms of two parameters $x_H$ and $y_O$, as sketched
in Fig.~\ref{fig:h2o}.
}
\label{tab:h2o}
\end{table}

\begin{figure}[ht!]
\includegraphics[width=0.45\textwidth]{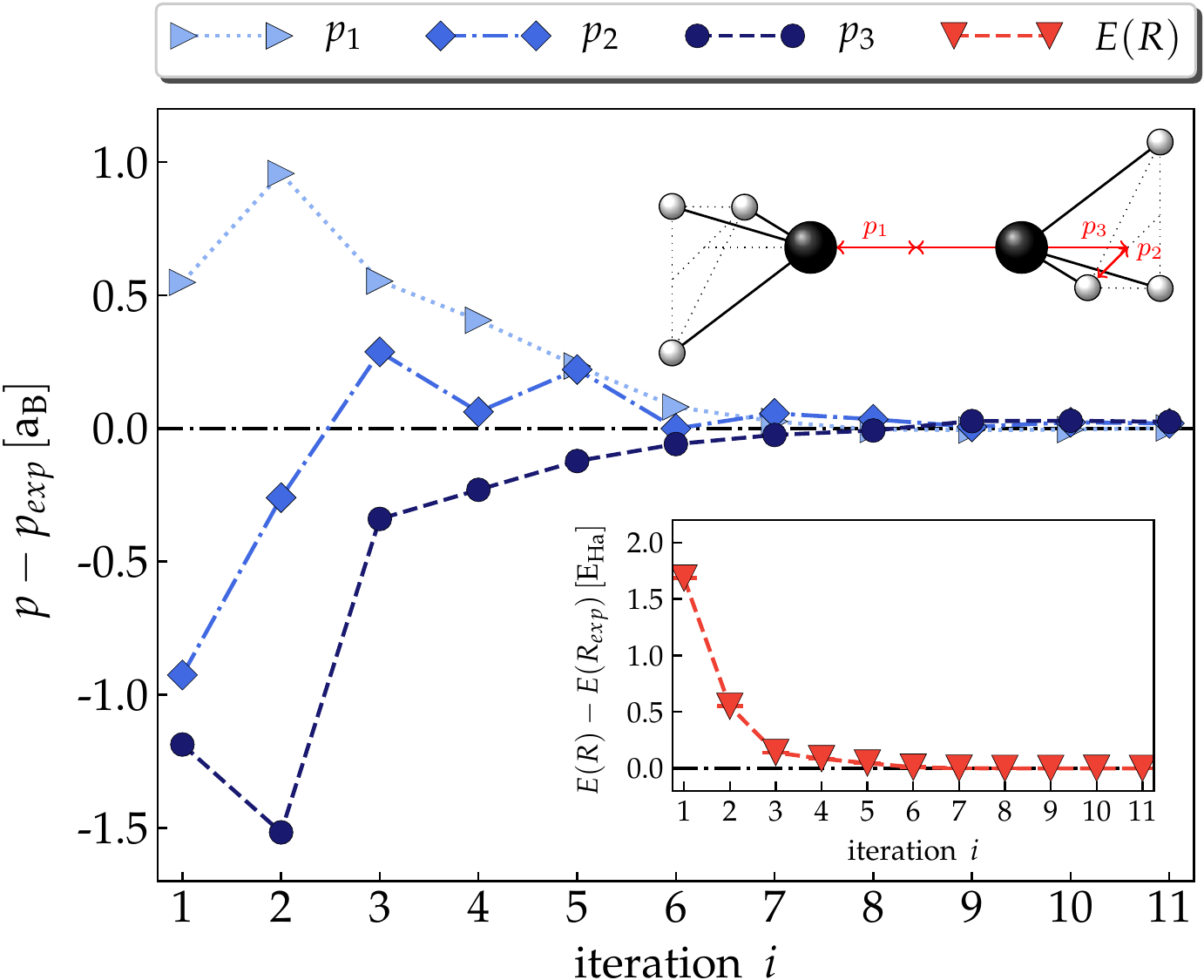}
\caption{(color online) Evolution of the molecular geometry for ethane during an AFQMC 
geometry optimization with the steepest descent algorithm. 
(The cc-pVDZ basis set was used here.)
The inset shows the corresponding total  energy.convergence.
Molecular geometries are expressed in terms of $3$ parameters, ${\bf{R}}={\bf{R}}({\bf{p}})$, 
as sketched in the upper right corner.
}
\label{fig:ethane}
\end{figure}

In Fig.~\ref{fig:ethane} we move to the more challenging case of ethane.
Geometries respecting the $\mathrm{D_3h}$ symmetry of the molecule can be expressed in terms of 
$3$ parameters, as sketched in the inset. 
As the number of geometric parameters grows, it quickly becomes impractical to compute the entire 
PES with AFQMC as was done in Fig.~\ref{fig:h2o}, and forced-based methods become essential.
The optimized parameters, with the modest but realistic cc-pVDZ basis, are 
within 0.02\,a.u.~(or 1\%)  of the experimental equilibrium geometry,
as seen in Table \ref{tab:ethane}.
The excellent agreement between the AFQMC geometry and those from high-level QC methods 
at this basis level suggests
that the discrepancy with experimental data is likely a result of basis set 
incompleteness, which is quantitatively confirmed by a cc-pVTZ calculation.
The ground-state energies computed at the optimized geometry and at the experimental equilibrium geometry
are in agreement with each other: $E({\bf R}_{\rm op})-E({\bf R}_{\rm eq})=-0.5(4)$\,mHa.

\begin{table}
\begin{tabular}{ccccc}
\hline\hline
basis & method & $p_1 \, [\mathrm{a_B}]$ & $p_2\, [\mathrm{a_B}]$ & $p_3\, [\mathrm{a_B}]$ \\
\hline
cc-pVDZ & DFT-B3LYP & 1.4447   & 1.9393   & 2.2080   \\
$ $     & QCISD(TQ) \cite{CCCBDB2016}     & 1.4513 & 1.9496 & 2.2079 \\
$ $     & CCSD(T)   & 1.4498   & 1.9476   & 2.2061   \\
$ $     & AFQMC     & 1.448(2) & 1.947(6) & 2.205(8) \\
cc-pVTZ & DFT-B3LYP & 1.4430   & 1.9202   & 2.1941   \\
$ $     & QCISD(TQ) \cite{CCCBDB2016}     & 1.4456 & 1.9271 & 2.1902 \\
$ $     & CCSD(T)   & 1.4391   & 1.9183   & 2.1803   \\
$ $     & AFQMC     & 1.443(5) & 1.912(8) & 2.192(6) \\
        & experiment \cite{CCCBDB2016}    & 1.4513 & 1.9260 & 2.1869 \\
\hline\hline
\end{tabular}
\caption{Equilibrium geometries of ethane from AFQMC. 
Results with the cc-pVDZ basis set are compared with 
several other QC methods.
Results from  cc-pVTZ indicate reasonable convergence with respect to basis set, and good 
agreement with experiment. 
DFT and CCSD(T) data were computed using the NWChem software \cite{Valiev_CPC_2010}.
Geometries are expressed in terms of the parameters $p_1$, $p_2$, $p_3$ sketched in Fig.~\ref{fig:ethane}.
}
\label{tab:ethane}
\end{table}

\begin{figure}
\includegraphics[width=0.45\textwidth]{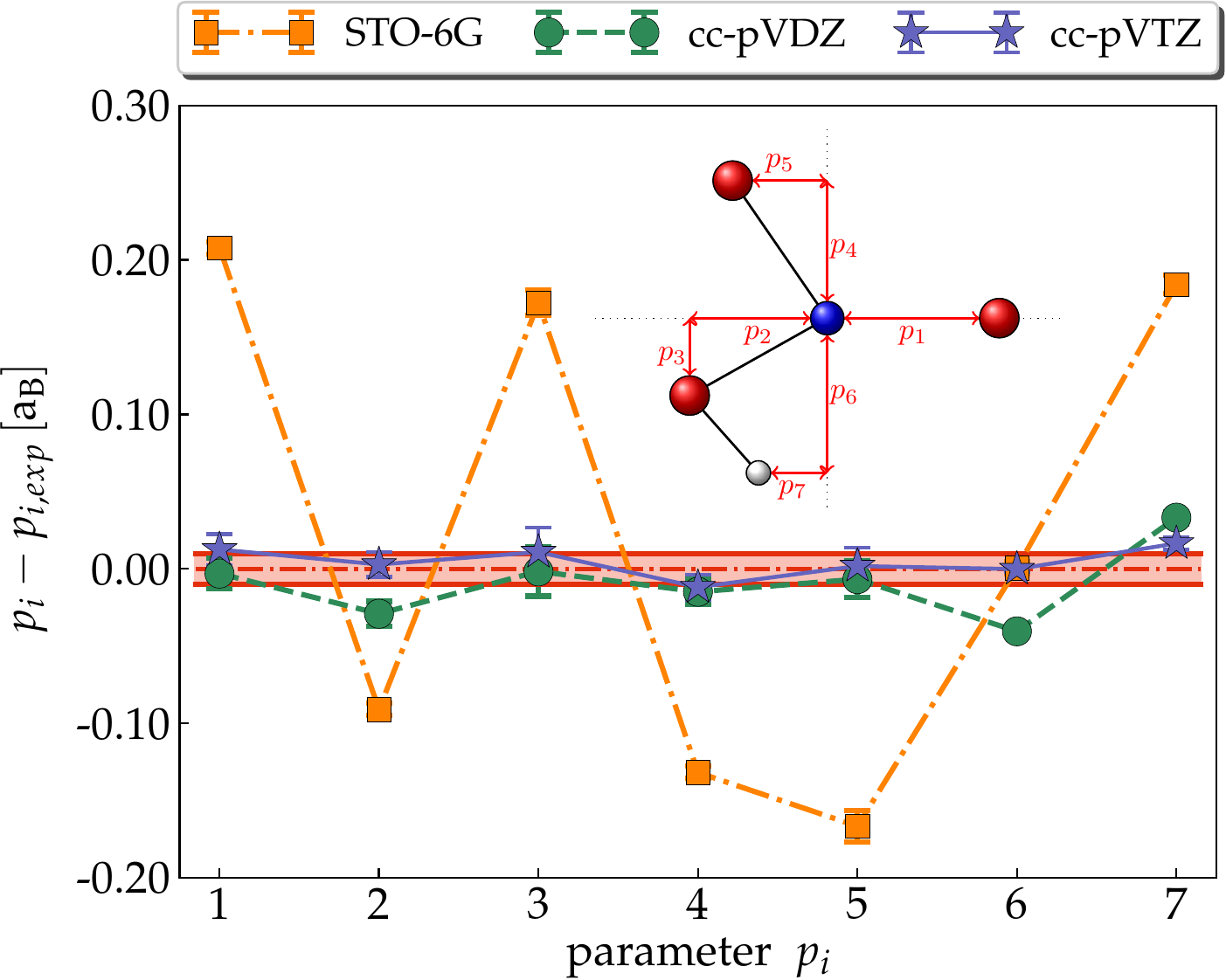}
\caption{(color online) Optimization of  nitric acid with AFQMC. 
The deviation between the final optimized geometry and the experimental equilibrium geometry
is shown for each basis set ( STO-6G, cc-pVDZ  and cc-pVTZ). 
The pink band indicates a range within 0.01\,a.u.~of experimental values.
}
\label{fig:nitric}
\end{figure}

As a last example, we optimize the molecular geometry of nitric acid.
We use increasingly larger basis sets to reach the continuum limit near the optimal geometry.
As shown in Fig.~\ref{fig:nitric}, there is a large residual error in the  STO-6G basis set, and 
the optimized geometry is significantly different from experiment. As more  realistic basis sets 
are employed, systematically improved results are obtained, and the computed geometry approaches 
the experimental equilibrium geometry. 
At the cc-pVTZ level, AFQMC results are in agreement with experiment 
to within 0.01 a.u.

We have demonstrated the direct computation of 
interatomic forces and molecular geometry optimization within AFQMC.
We proposed an internally consistent, numerically stable and computationally efficient
algorithm based on the Hellman-Feynman theorem and Pulay's corrections.
Results from a first application are presented. 
Accurate forces are obtained using simple RHF Slater determinant as constraining trial wave function. 
These results pave the way for systematic geometry optimization and potentially 
molecular dynamics using one of the most accurate many-body methods with low-power computational 
scaling. 

A variety of future directions are possible, including both applications and further 
 generalization and improvement of the algorithm. 
 More refined Ans\"atze \cite{Purwanto_JCP130_2009,Purwanto_JCP142_2015,Purwanto_JCP144_2016}
 for the trial wave function can be adopted straightforwardly for more challenging molecules. 
 Interfacing with better optimization strategies and improving the efficiency of the 
 force computation algorithm  with AFQMC itself can both lead to major increases in the capability of the
 method. 
 Alternative representations
\cite{Purwanto_JCP135_2011,Ma_PRL114_2015} of the Hamiltonian operator should be explored.
Applications to a variety of molecular systems are within reach including those containing 
 post-second-row elements. Geometry optimization in crystalline solids is being investigated.


We acknowledge support by NSF (Grant no. DMR-1409510), DOE (Grant no.~DE-SC0001303),
and the Simons Foundation. 
Computations were carried out at the Extreme Science and Engineering Discovery 
Environment (XSEDE), which is supported by National Science Foundation grant 
number ACI-1053575, at the Storm and SciClone Clusters at the College of 
William and Mary.
M. M. acknowledges James Shee and Qiming Sun for valuable interaction.

\appendix

\section{AFQMC estimator of interatomic forces}

The starting point for the evaluation of forces is the general form of the ground-state expectation value of the 
Hamiltonian operator,
\begin{equation}
\label{eq:app1}
E({\bf{R}}) = \frac
{\int d\Phi d\Psi \, \rho_{m,n}(\Phi,\Psi) \, W(\Phi,\Psi;{\bf{R}}) \, E_{loc}(\Phi,\Psi;{\bf{R}}) }
{\int d\Phi d\Psi \, \rho_{m,n}(\Phi,\Psi) \, W(\Phi,\Psi;{\bf{R}}) } \,.
\end{equation}
The Slater determinant 
$|\Psi \rangle = \prod_{\sigma} \prod_{l=1}^{N_\sigma} \psi_{l\sigma}^\dagger |\emptyset\rangle$ 
has $N_\sigma$ ($\sigma=\uparrow$ or $\downarrow$) fermions 
occupying orbitals 
$|\psi_{l\sigma} \rangle \equiv  \psi_{l\sigma}^\dagger |\emptyset\rangle=\sum_i \left( C_{\Psi,\sigma} \right)_{il} |\chi_i \rangle$, where $\{|\chi_i\rangle, i=1 \dots M\}$ are the basis functions and $C_{\Psi,\sigma} $ is a 
$M \times N_\sigma$ matrix that contains the orbital coefficients.
The Slater determinant $|\Phi \rangle$ is similarly defined in terms of orbitals $|\phi_{l\sigma}\rangle $ and 
matrix  $C_{\Phi,\sigma} $.
In Eq.~\eqref{eq:app1}, the overlap is given by 
\begin{equation}
\begin{split}
&W( \Phi , \Psi; {\bf{R}}) \equiv \langle \Phi | \Psi \rangle \\
= 
&\prod_\sigma \mbox{det} 
\left( C_{\Phi,\sigma}^\dagger S({\bf{R}}) C^{\phantom{\dagger}}_{\Psi, \sigma} \right) 
\equiv
\prod_\sigma \mbox{det} \left( \Omega_\sigma({\bf{R}}) \right)
\,,
\end{split}
\end{equation}
where $S({\bf{R}})_{ij} = \langle \chi_i | \chi_j \rangle$ is the overlap matrix of the AO basis set.
The local energy is given by
\begin{widetext}
\begin{equation}
\label{eq:app2}
\begin{split}
E_{loc}(\Phi,\Psi;{\bf{R}}) 
= H_0({\bf{R}}) + \sum_\sigma \mbox{Tr}[h({\bf{R}}) G_\sigma({\bf{R}})] + 
\sum_\gamma \Big( \sum_\sigma \mbox{Tr}[L^\gamma({\bf{R}}) G_\sigma({\bf{R}})] \Big)^2 
- 
\sum_\sigma \mbox{Tr}[L^\gamma({\bf{R}}) G_\sigma({\bf{R}})L^\gamma({\bf{R}}) G_\sigma({\bf{R}})]\,,
\end{split}
\end{equation}
\end{widetext}
where 
$H_0({\bf{R}})$ is a constant from the internuclear repulsion, 
the matrix $h({\bf{R}})$ is 
the one-body part of the Hamiltonian in the AO basis
($h({\bf{R}})_{ij} = \langle \chi_i | h_1({\bf{R}}) | \chi_j \rangle$), 
 the matrices $L^\gamma({\bf{R}})$ arise from decomposing  the two-body part of the Hamiltonian
 (e.g., via the Cholesky or density-fitting decomposition 
 \cite{Purwanto_JCP135_2011,Sun_arxiv_2017}, 
 $(ik|jl) = \sum_\gamma L^\gamma({\bf{R}})_{ik} L^\gamma({\bf{R}})_{jl}$),
and the (spin-dependent) one-electron Green's function is
\begin{equation}
\label{eq:app3}
G_\sigma({\bf{R}}) = 
C^{\phantom{\dagger}}_{\Psi, \sigma}
\Omega_\sigma({\bf{R}})^{-1}
C_{\Phi,\sigma}^\dagger\,,
\end{equation}
which obeys the idempotence relation
\begin{equation}
\label{eq:idempotence}
G_\sigma({\bf{R}}) = G_\sigma({\bf{R}}) S_\sigma({\bf{R}}) G_\sigma({\bf{R}}) \,.
\end{equation}

The gradient of $W$ can be obtained from
Jacobi's formula for the derivative of the determinant, 
\begin{equation}
\frac{\partial_{R_\alpha} W( \Phi , \Psi; {\bf{R}})}
{W( \Phi , \Psi; {\bf{R}})} = 
\sum_\sigma \mbox{Tr} \left( \partial_{R_ \alpha} S({\bf{R}}) \,G_\sigma({\bf{R}}) \right) \,.
\end{equation}
The gradient of $E_{loc}(\Phi,\Psi;{\bf{R}})$ is readily computed by recalling the cyclicity of the trace,
and that differentiation of \eqref{eq:app3} yields 
\begin{equation}
\partial_{R_\alpha} G_\sigma({\bf{R}}) = 
- G_\sigma({\bf{R}})\,\partial_{R_\alpha} S_\sigma({\bf{R}})\, G_\sigma({\bf{R}}) \,,
\end{equation}
complying with the idempotence relation \eqref{eq:idempotence} \cite{Pulay_MP17_1969}. The gradient 
$\partial_{R_\alpha} G_\sigma({\bf{R}})$ incorporates Pulay's corrections in the framework
of AFQMC, which are important for an accurate evaluation of forces when atom-centered Gaussian
orbitals are used as basis for the one-electron Hilbert space.
The presence of Pulay's corrections is reminiscent of the covariant derivatives that appear in fiber bundle
theory and differential geometry \cite{Helgaker_chap_1986}. This reminiscence is not accidental: the 
purpose of corrective terms in covariant derivatives is to preserve some properties of the space on which
they act. For example, in gauge theories corrections preserve the gradient of the wavefunction under 
gauge transformations and, on Riemann manifolds, connections preserve parallelism.
In this framework, Pulay's correction preserve the idempotency \eqref{eq:idempotence} of the one-particle
reduced Green's function under changes in molecular geometries.

The gradient of $E_{loc}(\Phi,\Psi;{\bf{R}})$ thus reads
\begin{widetext}
\begin{equation}
\label{eq:gradient_detail}
\begin{split}
&\partial_{R_\alpha} E_{loc}(\Phi,\Psi;{\bf{R}}) 
= \partial_{R_\alpha} H_0({\bf{R}}) + \sum_\sigma \mbox{Tr}[\partial_{R_\alpha} h({\bf{R}})\,G_\sigma({\bf{R}})] + \sum_\sigma \mbox{Tr}[h({\bf{R}})\,\partial_{R_\alpha} G_\sigma({\bf{R}})] \\
+ & 2\,\sum_{\gamma,\sigma{\sigma'}} \mbox{Tr}[L^\gamma({\bf{R}})G_\sigma({\bf{R}})]\, 
 \mbox{Tr}[\partial_{R_\alpha} L^\gamma({\bf{R}})\,G_{\sigma'}({\bf{R}})]  
- 2\,\sum_{\gamma,\sigma} \mbox{Tr}[\partial_{R_\alpha} L^\gamma({\bf{R}})\, G_\sigma({\bf{R}})L^\gamma({\bf{R}}) G_\sigma({\bf{R}})] + \\
+ & 2\,\sum_{\gamma,\sigma{\sigma'}} \mbox{Tr}[L^\gamma({\bf{R}}) G_\sigma({\bf{R}})]\, 
  \mbox{Tr}[ L^\gamma({\bf{R}})\,\partial_{R_\alpha} G_{\sigma'}({\bf{R}})]  
- 2 \sum_{\gamma,\sigma} \mbox{Tr}[L^\gamma({\bf{R}}) G_\sigma({\bf{R}})L^\gamma({\bf{R}})\, \partial_{R_\alpha} G_\sigma({\bf{R}})]\,.
\end{split}
\end{equation}
\end{widetext}
Computing the local energy requires $\mathcal{O}(N_\gamma (M^3 + M + M^2))$ operations, 
while computing all 
components of the force requires $\mathcal{O}(N_\gamma (3 M^3 + M + M^2 + 4 N_{R_\alpha} M^2))$ operations, 
with $N_{R_\alpha}$ being 
the number of force 
 components. 
 Thus the cost for computing all force components is about a factor 
 $(3+4\frac{N_p}{M})$ of that of the local energy. The ratio is of course bounded by $7$ and should
 be approximately $3$ in most situations, which was confirmed in our studies. 
t should be possible to exploit the structure of the components of $L$ and possibly regroup
terms in Eq.~\eqref{eq:gradient_detail} to further speed up the computation.
The AFQMC estimator of the force is then given by
\begin{widetext}
\begin{equation}
F^\alpha({\bf{R}}) = 
\frac
{\int d\Phi d\Psi \, \rho_{m,n}(\Phi,\Psi) \, W(\Phi,\Psi;{\bf{R}}) \, \left( \partial_{R_\alpha} E_{loc}(\Phi,\Psi;{\bf{R}}) + 
\frac{\partial_{R_\alpha} W( \Phi , \Psi; {\bf{R}}) }{W( \Phi , \Psi; {\bf{R}})} \big(E_{loc}(\Phi,\Psi;{\bf{R}}) - E({\bf{R}}) \big) \right) }
{\int d\Phi d\Psi \, \rho_{m,n}(\Phi,\Psi) \, W(\Phi,\Psi;{\bf{R}})}\,.
\end{equation}
\end{widetext}
When $m=n=0$, 
$\rho_{m,n}(\Phi,\Psi) = \delta(\Phi-\Psi_T) \delta(\Psi-\Psi_T)$,
so that $E({\bf{R}}) = E_{loc}(\Psi_T,\Psi_T;{\bf{R}})$ and the estimator reduces 
to the familiar Hartree-Fock expression \cite{Pulay_MP17_1969}.

In this work, the derivatives $\partial_{R_\alpha} S$, $\partial_{R_\alpha} h$ and $\partial_{R_\alpha} 
L^\gamma$ were all obtained by numerical differentiation. A finite difference step size of $10^{-4}$ a.u.~was used, and it was verified
that using a smaller step does not change the results within statistical error bars.


\begin{thebibliography}{10}

\bibitem{Car_PRL55_1985}
R.~Car and M.~Parrinello, ``Unified approach for molecular dynamics and
  density-functional theory,'' {\em Phys. Rev. Lett.}, vol.~55, pp.~2471--2474,
  Nov 1985.

\bibitem{Pulay_MP17_1969}
P.~Pulay, ``Analytical derivatives, forces, force constants, molecular
  geometries, and related response properties in electronic structure theory,''
  {\em Wiley Interdisciplinary Reviews: Computational Molecular Science},
  vol.~4, no.~3, pp.~169--181, 2014.

\bibitem{Eckert_JCC18_1997}
F.~Eckert, P.~Pulay, and H.-J. Werner, ``Ab initio geometry optimization for
  large molecules,'' {\em Journal of Computational Chemistry}, vol.~18, no.~12,
  pp.~1473--1483, 1997.

\bibitem{Pulay_WIRES4_2014}
P.~Pulay, ``Ab initio calculation of force constants and equilibrium geometries
  in polyatomic molecules,'' {\em Molecular Physics}, vol.~17, no.~2,
  pp.~197--204, 1969.

\bibitem{Eckert_TCA100_1998}
F.~Eckert and H.-J. Werner, ``Reaction path following by quadratic steepest
  descent,'' {\em Theoretical Chemistry Accounts}, vol.~100, pp.~21--30, Nov
  1998.

\bibitem{Reveles_JCC25_2004}
J.~U. Reveles and A.~M. Koster, ``Geometry optimization in density functional
  methods,'' {\em J. Comp. Chem.}, vol.~25, no.~9, pp.~1109--1116, 2004.

\bibitem{Busch_JCP94_1991}
T.~Busch, A.~D. Esposti, and H.-J. Werner, ``Analytical energy gradients for
  multiconfiguration self-consistent field wave functions with frozen core
  orbitals,'' {\em J. Chem. Phys.}, vol.~94, no.~10, pp.~6708--6715, 1991.

\bibitem{Azhary_JCP108_1998}
A.~E. Azhary, G.~Rauhut, P.~Pulay, and H.-J. Werner, ``Analytical energy
  gradients for local second-order {M}\"oller-{P}lesset perturbation theory,''
  {\em J. Chem. Phys.}, vol.~108, no.~13, pp.~5185--5193, 1998.

\bibitem{Rauhut_PCCP3_2001}
G.~Rauhut and H.-J. Werner, ``Analytical energy gradients for local
  coupled-cluster methods,'' {\em Phys. Chem. Chem. Phys.}, vol.~3,
  pp.~4853--4862, 2001.

\bibitem{Zhang_PRL90_2003}
S.~Zhang and H.~Krakauer, ``Quantum {M}onte {C}arlo method using phase-free
  random walks with slater determinants,'' {\em Phys. Rev. Lett.}, vol.~90,
  p.~136401, Apr 2003.

\bibitem{AlSaidi_JCP124_2006}
W.~A. Al-Saidi, S.~Zhang, and H.~Krakauer, ``Auxiliary-field quantum {M}onte
  {C}arlo calculations of molecular systems with a {G}aussian basis,'' {\em J.
  Chem. Phys.}, vol.~124, no.~22, p.~224101, 2006.

\bibitem{Suewattana_PRB75_2007}
M.~Suewattana, W.~Purwanto, S.~Zhang, H.~Krakauer, and E.~J. Walter,
  ``Phaseless auxiliary-field quantum {M}onte {C}arlo calculations with plane
  waves and pseudopotentials: Applications to atoms and molecules,'' {\em Phys.
  Rev. B}, vol.~75, p.~245123, Jun 2007.

\bibitem{Purwanto_JCP128_2008}
W.~Purwanto, W.~A. Al-Saidi, H.~Krakauer, and S.~Zhang, ``Eliminating spin
  contamination in auxiliary-field quantum {M}onte {C}arlo: Realistic potential
  energy curve of ${F}_2$,'' {\em J. Chem. Phys.}, vol.~128, no.~11, p.~114309,
  2008.

\bibitem{Purwanto_JCP130_2009}
W.~Purwanto, S.~Zhang, and H.~Krakauer, ``Excited state calculations using
  phaseless auxiliary-field quantum {M}onte {C}arlo: Potential energy curves of
  low-lying ${C}_2$ singlet states,'' {\em J. Chem. Phys.}, vol.~130, no.~9,
  p.~094107, 2009.

\bibitem{Purwanto_JCP142_2015}
W.~Purwanto, S.~Zhang, and H.~Krakauer, ``An auxiliary-field quantum {M}onte
  {C}arlo study of the chromium dimer,'' {\em J. Chem. Phys.}, vol.~142, no.~6,
  p.~064302, 2015.

\bibitem{Purwanto_JCP144_2016}
W.~Purwanto, S.~Zhang, and H.~Krakauer, ``Auxiliary-field quantum {M}onte
  {C}arlo calculations of the molybdenum dimer,'' {\em J. Chem. Phys.},
  vol.~144, no.~24, p.~244306, 2016.

\bibitem{Kwee_PRL100_2008}
H.~Kwee, S.~Zhang, and H.~Krakauer, ``Finite-size correction in many-body
  electronic structure calculations,'' {\em Phys. Rev. Lett.}, vol.~100,
  p.~126404, Mar 2008.

\bibitem{Purwanto_PRB80_2009}
W.~Purwanto, H.~Krakauer, and S.~Zhang, ``Pressure-induced diamond to
  $\ensuremath{\beta}$-tin transition in bulk silicon: A quantum {M}onte
  {C}arlo study,'' {\em Phys. Rev. B}, vol.~80, p.~214116, Dec 2009.

\bibitem{Ma_PRL114_2015}
F.~Ma, W.~Purwanto, S.~Zhang, and H.~Krakauer, ``Quantum {M}onte {C}arlo
  calculations in solids with downfolded hamiltonians,'' {\em Phys. Rev.
  Lett.}, vol.~114, p.~226401, Jun 2015.

\bibitem{Shee2017}
J.~Shee, S.~Zhang, D.~R. Reichman, and R.~A. Friesner, ``Chemical
  transformations approaching chemical accuracy via correlated sampling in
  auxiliary-field quantum {M}onte {C}arlo,'' {\em J. Chem. Theor. Comput.},
  vol.~13, no.~6, pp.~2667--2680, 2017.
\newblock PMID: 28481546.

\bibitem{Motta_PRX_2017}
M.~Motta, D.~M. Ceperley, G.~K.-L. Chan, J.~A. Gomez, E.~Gull, S.~Guo, C.~A.
  Jim\'enez-Hoyos, T.~N. Lan, J.~Li, F.~Ma, A.~J. Millis, N.~V. Prokof'ev,
  U.~Ray, G.~E. Scuseria, S.~Sorella, E.~M. Stoudenmire, Q.~Sun, I.~S.
  Tupitsyn, S.~R. White, D.~Zgid, and S.~Zhang, ``Towards the solution of the
  many-electron problem in real materials: Equation of state of the hydrogen
  chain with state-of-the-art many-body methods,'' {\em Phys. Rev. X}, vol.~7,
  p.~031059, Sep 2017.

\bibitem{Esler_JP125_2008}
K.~P. Esler, J.~Kim, D.~M. Ceperley, W.~Purwanto, E.~J. Walter, H.~Krakauer,
  S.~Zhang, P.~R.~C. Kent, R.~G. Hennig, C.~Umrigar, M.~Bajdich, J.~Kolorenč,
  L.~Mitas, and A.~Srinivasan, ``Quantum {M}onte {C}arlo algorithms for
  electronic structure at the petascale; the {E}ndstation project,'' {\em
  Journal of Physics: Conference Series}, vol.~125, no.~1, p.~012057, 2008.

\bibitem{Zong_PRE58_1998}
F.~Zong and D.~M. Ceperley, ``Path integral {M}onte {C}arlo calculation of
  electronic forces,'' {\em Phys. Rev. E}, vol.~58, pp.~5123--5130, Oct 1998.

\bibitem{Casalegno_JCP118_2003}
M.~Casalegno, M.~Mella, and A.~M. Rappe, ``Computing accurate forces in quantum
  {M}onte {C}arlo using {P}ulay's corrections and energy minimization,'' {\em
  J. Chem. Phys.}, vol.~118, no.~16, pp.~7193--7201, 2003.

\bibitem{Lee_JCP122_2005}
M.~W. Lee, M.~Mella, and A.~M. Rappe, ``Electronic quantum {M}onte {C}arlo
  calculations of atomic forces, vibrations, and anharmonicities,'' {\em J.
  Chem. Phys.}, vol.~122, no.~24, p.~244103, 2005.

\bibitem{Sorella_JCP133_2010}
S.~Sorella and L.~Capriotti, ``Algorithmic differentiation and the calculation
  of forces by quantum {M}onte {C}arlo,'' {\em J. Chem. Phys.}, vol.~133,
  no.~23, p.~234111, 2010.

\bibitem{Assaraf_JCP113_2000}
R.~Assaraf and M.~Caffarel, ``Computing forces with quantum {M}onte {C}arlo,''
  {\em J. Chem. Phys.}, vol.~113, no.~10, pp.~4028--4034, 2000.

\bibitem{Assaraf_JCP119_2003}
R.~Assaraf and M.~Caffarel, ``Zero-variance zero-bias principle for observables
  in quantum {M}onte {C}arlo: Application to forces,'' {\em J. Chem. Phys.},
  vol.~119, no.~20, pp.~10536--10552, 2003.

\bibitem{Filippi_PRB61_2000}
C.~Filippi and C.~J. Umrigar, ``Correlated sampling in quantum {M}onte {C}arlo:
  A route to forces,'' {\em Phys. Rev. B}, vol.~61, pp.~R16291--R16294, Jun
  2000.

\bibitem{Chiesa_PRL94_2005}
S.~Chiesa, D.~M. Ceperley, and S.~Zhang, ``Accurate, efficient, and simple
  forces computed with quantum {M}onte {C}arlo methods,'' {\em Phys. Rev.
  Lett.}, vol.~94, p.~036404, Jan 2005.

\bibitem{Moroni_JCTC10_2014}
S.~Moroni, S.~Saccani, and C.~Filippi, ``Practical schemes for accurate forces
  in quantum {M}onte {C}arlo,'' {\em J. Chem. Theor. Comput.}, vol.~10, no.~11,
  pp.~4823--4829, 2014.
\newblock PMID: 26584369.

\bibitem{Motta_JCTC_2017}
M.~Motta and S.~Zhang, ``Computation of ground-state properties in molecular
  systems: back-propagation with auxiliary-field quantum {M}onte {C}arlo,''
  {\em arXiv:1707.02684}, 2017.

\bibitem{Zhang_notes_2013}
S.~Zhang, ``Auxiliary-field quantum {M}onte {C}arlo for correlated electron
  systems,'' in {\em Emergent Phenomena in Correlated Matter: Modeling and
  Simulation} (E.~P.~E. Koch and U.~Schollw\"ock, eds.), ch.~15, Verlag des
  Forschungszentrum J\"ulich, 2013.

\bibitem{Motta_WIRES_2017}
M.~Motta and S.~Zhang, ``Ab initio computations of molecular systems by the
  auxiliary-field quantum {M}onte {C}arlo method,'' {\em arXiv:1711.02242},
  2017.

\bibitem{Fahy_PRB43_1991}
S.~Fahy and D.~R. Hamann, ``Diffusive behavior of states in the
  hubbard-stratonovitch transformation,'' {\em Phys. Rev. B}, vol.~43,
  pp.~765--779, Jan 1991.

\bibitem{Purwanto_PRE70_2004}
W.~Purwanto and S.~Zhang, ``Quantum {M}onte {C}arlo method for the ground state
  of many-boson systems,'' {\em Phys. Rev. E}, vol.~70, p.~056702, Nov 2004.

\bibitem{LeBlanc_PRX5_2015}
J.~P.~F. LeBlanc, A.~E. Antipov, F.~Becca, I.~W. Bulik, G.~K.-L. Chan, C.-M.
  Chung, Y.~Deng, M.~Ferrero, T.~M. Henderson, C.~A. Jim\'enez-Hoyos, E.~Kozik,
  X.-W. Liu, A.~J. Millis, N.~V. Prokof'ev, M.~Qin, G.~E. Scuseria, H.~Shi,
  B.~V. Svistunov, L.~F. Tocchio, I.~S. Tupitsyn, S.~R. White, S.~Zhang, B.-X.
  Zheng, Z.~Zhu, and E.~Gull, ``Solutions of the two-dimensional hubbard model:
  Benchmarks and results from a wide range of numerical algorithms,'' {\em
  Phys. Rev. X}, vol.~5, p.~041041, Dec 2015.

\bibitem{Hellmann_book_1937}
H.~Hellmann, {\em Einf{\"u}hrung in die Quantenchemie}.
\newblock Franz Deuticke, 1937.

\bibitem{Feynman_PR56_1939}
R.~P. Feynman, ``Forces in molecules,'' {\em Phys. Rev.}, vol.~56,
  pp.~340--343, Aug 1939.

\bibitem{Zhang_PRB_1997}
S.~Zhang, J.~Carlson, and J.~E. Gubernatis, ``Constrained path {M}onte {C}arlo
  method for fermion ground states,'' {\em Phys. Rev. B}, vol.~55,
  pp.~7464--7477, Mar 1997.

\bibitem{CCCBDB2016}
R.~D.~J. III, {\em NIST Computational Chemistry Comparison and Benchmark
  Database (CCC-BDB)}.
\newblock NIST Standard Reference Database Number 101, Release 18, 2016.

\bibitem{Valiev_CPC_2010}
M.~Valiev, E.~Bylaska, N.~Govind, K.~Kowalski, T.~Straatsma, H.~V. Dam,
  D.~Wang, J.~Nieplocha, E.~Apra, T.~Windus, and W.~de~Jong, ``Nwchem: A
  comprehensive and scalable open-source solution for large scale molecular
  simulations,'' {\em Computer Physics Communications}, vol.~181, no.~9,
  pp.~1477 -- 1489, 2010.

\bibitem{Purwanto_JCP135_2011}
W.~Purwanto, H.~Krakauer, Y.~Virgus, and S.~Zhang, ``Assessing weak hydrogen
  binding on {C}a centers: an accurate many-body study with large basis sets,''
  {\em J. Chem. Phys.}, vol.~135, no.~16, p.~164105, 2011.

\bibitem{Sun_arxiv_2017}
Q.~Sun, T.~C. Berkelbach, N.~S. Blunt, G.~H. Booth, S.~Guo, Z.~Li, J.~Liu,
  J.~McClain, E.~R.Sayfutyarova, S.~Sharma, S.~Wouters, and G.~K.-L. Chan,
  ``The python-based simulations of chemistry framework (pyscf),'' {\em WIREs
  Comput. Mol. Sci.}

\bibitem{Helgaker_chap_1986}
T.~U. Helgaker, ``Hamiltonian expansion in geometrical distortions,'' in {\em
  Geometrical Derivatives of Energy Surfaces and Molecular Properties} (J.~S.
  Poul~Jorgensen, ed.), ch.~1, NATO Science Series C, 1986.

\end{thebibliography}
\end{document}